\documentclass[aps,twocolumn,amsmath,amssymb,showpacs,prl,superscriptaddress,unsortedaddress]{revtex4}
\usepackage{epsf}
\usepackage{graphicx}
\usepackage{gensymb}

\newcommand{\etal}{{\it et al.}}

\begin{document}

\title{The Fermi surface of Ba$_{1-x}$K$_x$Fe$_2$As$_2$ and its evolution with doping}

\author{C. Liu}
\affiliation{Ames Laboratory and Department of Physics and
Astronomy, Iowa State University, Ames, Iowa 50011, USA}

\author{G. D. Samolyuk}
\affiliation{Ames Laboratory and Department of Physics and
Astronomy, Iowa State University, Ames, Iowa 50011, USA}

\author{Y. Lee}
\affiliation{Ames Laboratory and Department of Physics and
Astronomy, Iowa State University, Ames, Iowa 50011, USA}

\author{N. Ni}
\affiliation{Ames Laboratory and Department of Physics and
Astronomy, Iowa State University, Ames, Iowa 50011, USA}

\author{T. Kondo}
\affiliation{Ames Laboratory and Department of Physics and
Astronomy, Iowa State University, Ames, Iowa 50011, USA}

\author{A. F. Santander-Syro}
\affiliation{Laboratoire Photons Et Mati\`ere, UPR-5 CNRS, ESPCI, 10 rue Vauquelin, 75231 Paris cedex 5, France}
 \affiliation{Labratoire de Physique des Solides, UMR-8502 CNRS, Universit\'e Paris-Sud, B\^at. 510, 91405 Orsay, France}

\author{S. L. Bud'ko}
\affiliation{Ames Laboratory and Department of Physics and
Astronomy, Iowa State University, Ames, Iowa 50011, USA}

\author{J. L. McChesney}
\affiliation{Advanced Light Source, Berkeley National Laboratory, Berkeley, California 94720, USA}

\author{E. Rotenberg}
\affiliation{Advanced Light Source, Berkeley National Laboratory, Berkeley, California 94720, USA}

\author{T. Valla}
\affiliation{Condensed Matter Physics and Materials Science Dept.,
Brookhaven National Laboratory, Upton, New York 11973, USA}

\author{A. V. Fedorov}
\affiliation{Advanced Light Source, Berkeley National Laboratory, Berkeley, California 94720, USA}

\author{P. C. Canfield}
\affiliation{Ames Laboratory and Department of Physics and
Astronomy, Iowa State University, Ames, Iowa 50011, USA}

\author{B. N. Harmon}
\affiliation{Ames Laboratory and Department of Physics and
Astronomy, Iowa State University, Ames, Iowa 50011, USA}

\author{A. Kaminski}
\affiliation{Ames Laboratory and Department of Physics and
Astronomy, Iowa State University, Ames, Iowa 50011, USA}

\date{\today}
\begin{abstract}
We use angle-resolved photoemission spectroscopy (ARPES) to
investigate the electronic properties of the newly discovered
iron-arsenic superconductor, Ba$_{1-x}$K$_x$Fe$_2$As$_2$ and
non-supercondcuting BaFe$_2$As$_2$. Our study indicates that the
Fermi surface of the undoped, parent compound BaFe$_2$As$_2$
consists of hole pocket(s) at $\Gamma$ (0,0) and larger electron
pocket(s) at X (1,0), in general agreement with full-potential
linearized plane wave (FLAPW) calculations. Upon doping with
potassium, the hole pocket expands and the electron pocket becomes
smaller with its bottom approaching the chemical potential. Such an
evolution of the Fermi surface is consistent with hole doping within
a rigid band shift model. Our results also indicate that FLAPW
calculation is a reasonable approach for modeling the electronic
properties of both undoped and K-doped iron arsenites.
\end{abstract}


\maketitle

Iron-Arsenic based materials comprise a very interesting class of
materials with many unusual properties. For example, they have
recently been shown to be superconducting with a $T_c$ as high as
55K\cite{Kamihara_original, Takahashi43K, Ren_55K}. This discovery
has initiated a frenzy of research activity, which until very
recently was limited to studies of only polycrystalline samples.
Initial experiments focused on fluorine-doped rare earth oxide based
materials (RFeAsOF) \cite{Chen,la_Cruz_NeutronScattering_SDW, Hunte}. 
To date, there is very little photoemission
data available on these compounds\cite{Sato_GapPE, Jia_PE, Liu_PE}
with only one angle resolved study\cite{CHANGLIU}. The recent
discovery of superconductivity in oxygen free
Ba$_{1-x}$K$_x$Fe$_2$As$_2$ \cite{Rotter} suggests that the
superconductivity is ultimately linked to the electronic properties
of the iron arsenic layer(s) with the remaining layers acting as a
charge reservoir. This scheme closely resembles the situation found
in the cuprates but without oxygen. In both  BaFe$_2$As$_2$ and
SrFe$_2$As$_2$ there are clear structural phase transitions from a
high temperature tetragonal to low temperature orthorhombic
phases.\cite{NiNi, Yan} When potassium is substituted for barium,
the temperature at which the structural transition occurs is
suppressed and superconductivity emerges\cite{Rotter, NiNi}. Some
experiments also point to the existence of a transition into a spin
density wave (SDW) state at higher temperature
\cite{la_Cruz_NeutronScattering_SDW, WANG_SDW} and related changes
of the electronic structure\cite{FENG}. Determining the effects of
doping on low lying electronic excitations is essential for this
study, as they play significant role in determining the normal state
and superconducting properties. It is equally important to
understand the electronic properties of the parent compound because
undoped systems are easier to model theoretically and they represent
a basis for higher order approximations. The information about
electronic structure and its evolution with doping is deemed
essential to formulate a successful model of superconductivity in
these fascinating systems.

The recent growth of large, high quality single crystals \cite{NiNi}
has opened up the possibility of examining the electronic properties
of these materials. Here we present data from angle resolved
photoemission spectroscopy (ARPES) on the Fermi surface and band
dispersion, and discuss how they are affected by doping with
potassium. We find in the undoped samples, the Fermi surface
consists of a smaller hole pocket centered at $\Gamma$ (0,0,0) and a
larger electron pocket located at each of the X points. Upon doping
with potassium, the $\Gamma$ hole pocket increases in size and the X
pocket contracts. There are also significant shifts in energy for
lower lying, fully occupied bands. In potassium doped system - on
the verge of superconductivity the bottom of X band is located in
close proximity to the chemical potential. This may have
significance for the emergence of the superconductivity in this
system. Our experimental results are in general agreement with band
calculations for both undoped and doped system.

\begin{figure*}
\includegraphics[width=5.5in]{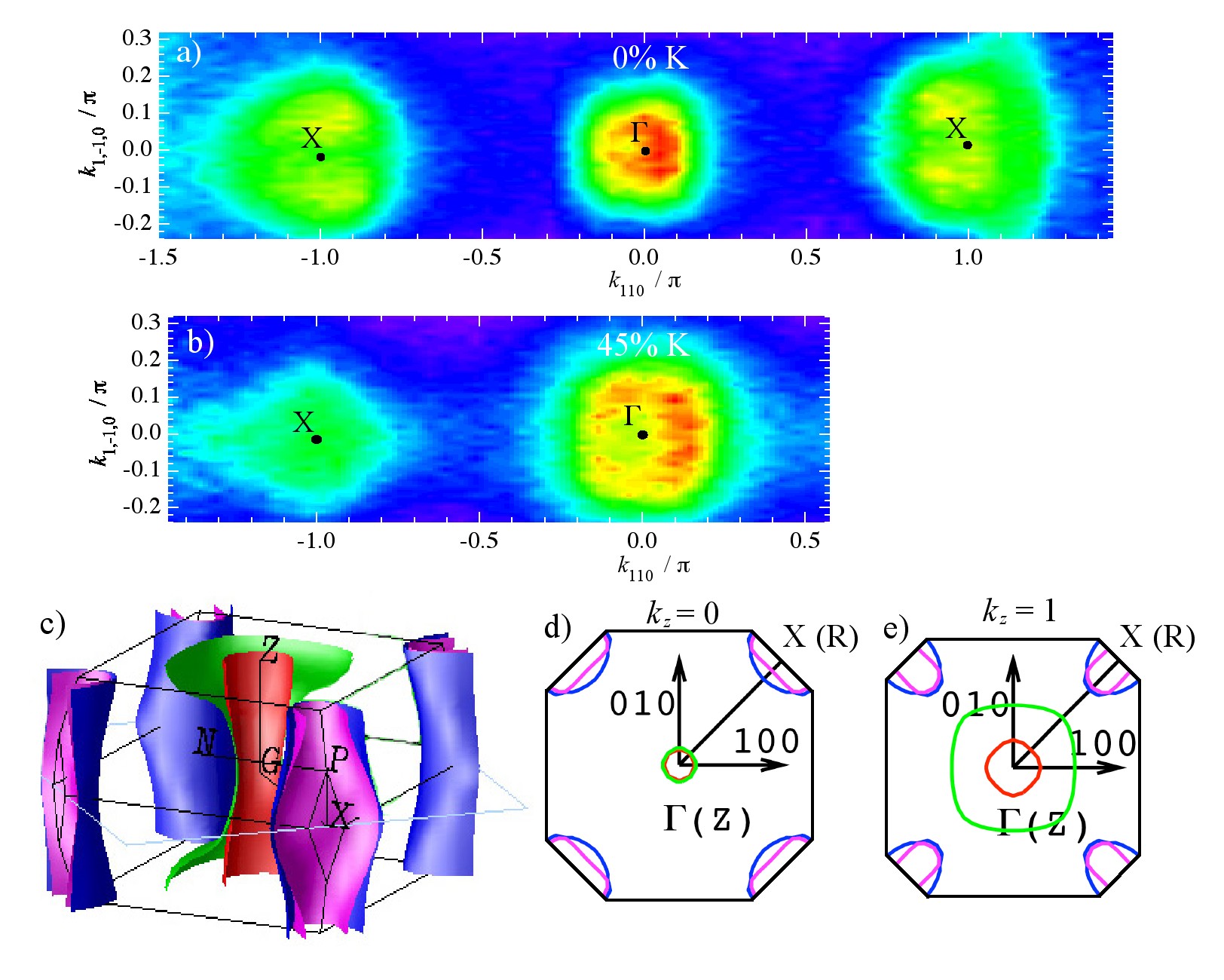}
\caption{(color online) Measured Fermi surface (FS) of
BaFe$_2$As$_2$ and Ba$_{1-x}$K$_x$Fe$_2$As$_2$ and calculated FS for
undoped case. a) FS map of BaFe$_2$As$_2$  - intensity of the
photoelectrons integrated over 20 meV about the chemical potential
obtained with 40.8 eV photons. Experiment was done at \textit{T} =
100K. Areas of bright color mark the locations of the FS. b) FS map
of Ba$_{1-x}$K$_x$Fe$_2$As$_2$ with nominal $x = 0.45$ measured
under the same conditions as (a). c) 3-dimensional FS of
BaFe$_2$As$_2$ obtained from FLAPW calculations. d) FS cross-section
for $k_{z} = 0$ (X-$\Gamma$ plane) obtained by FLAPW calculations.
e) same as (d) but for $k_{z} = 1$ (R-Z plane). } \label{fig1}
\end{figure*}

%%%%%%%%%%
%%%%%
%Experimental (will be moved to methods section for NP)
%%%%%%%%%%%%%%%%%%%%%%
%samples
Single crystals of the parent compound, BaFe$_2$As$_2$, and potassium doped
samples, Ba$_{1-x}$K$_x$Fe$_2$As$_2$, with an approximate doping $x =
0.45$ were grown out of a Sn flux using conventional high
temperature solution growth techniques\cite{NiNi}. Large (up to 2$\times$2
mm) single crystals were cleaved \textit{in situ} yielding flat
mirror-like surfaces.
%ARPES
The experimental data was acquired using a laboratory-based ARPES
system consisting of a Scienta SES2002 electron analyzer, GammaData
UV lamp and custom designed refocusing optics. The samples were cooled
using a closed-cycle refrigerator. Measurements were performed on
several samples and all yielded similar results for the band dispersion
and Fermi surface. All data were acquired using the HeII line with
a photon energy of 40.8 eV. The momentum resolution was set at 0.014
\AA$^{-1}$ and 0.06 \AA$^{-1}$ parallel and perpendicular to the
slit direction. The energy resolution was 30 meV for the Fermi
surface scans and 15 meV for the intensity maps.
%Calculations
Our full-potential linearized plane wave (FLAPW) calculation \cite{FLAPW}  have done with local density approximation(LDA) \cite{LDA} , and the experimental lattice constants \cite{Rotter} for the undoped parent compound BaFe$_2$As$_2$ and potassium doped Ba$_{1-x}$K$_x$Fe$_2$As$_2$ with $x = 0.4$. Total energy minimization was used to determine $z_{\textrm{As}}$ = 0.341c. 

%%%%END OF EXPERIMENTAL SECTION

\begin{figure}
\includegraphics[width=3.4in]{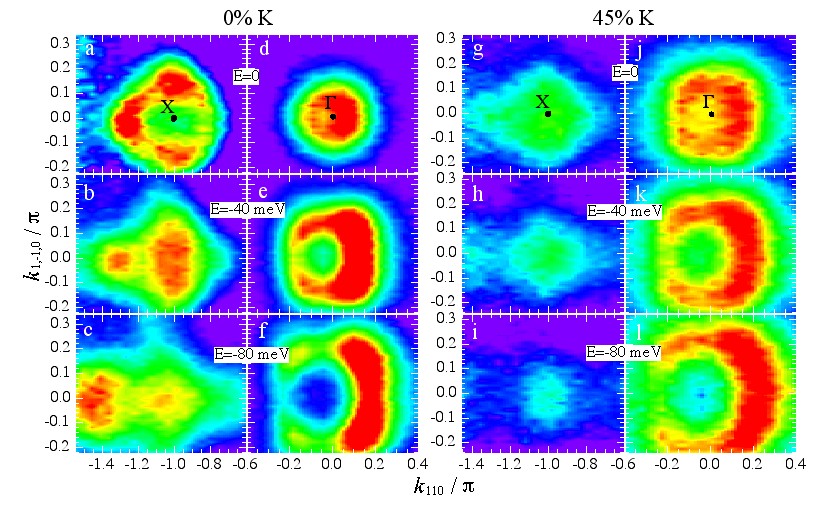}
\caption{(color online) Electron intensity map for selected binding
energies of the undoped and potassium-doped samples. All data was
taken at \textit{T} = 100K. Panels (a)-(f) present the data of
undoped BaFe$_2$As$_2$, while panels (g)-(l) present the data of
Ba$_{1-x}$K$_x$Fe$_2$As$_2$ with nominal $x = 0.45$. The three rows
of panels present the data at the chemical potential
\textit{E}$_{\textrm{F}}$, binding energy \textit{E} = 0.04 and
0.08eV, respectively; the two columns of panels for each doping
present the intensity map covering a \textit{k}-space area near the
X-point [(a)-(c), (g)-(i)] and the $\Gamma$-point [(d)-(f),
(j)-(l)], respectively. } \label{fig2}
\end{figure}

\begin{figure*}
\includegraphics[width=6in]{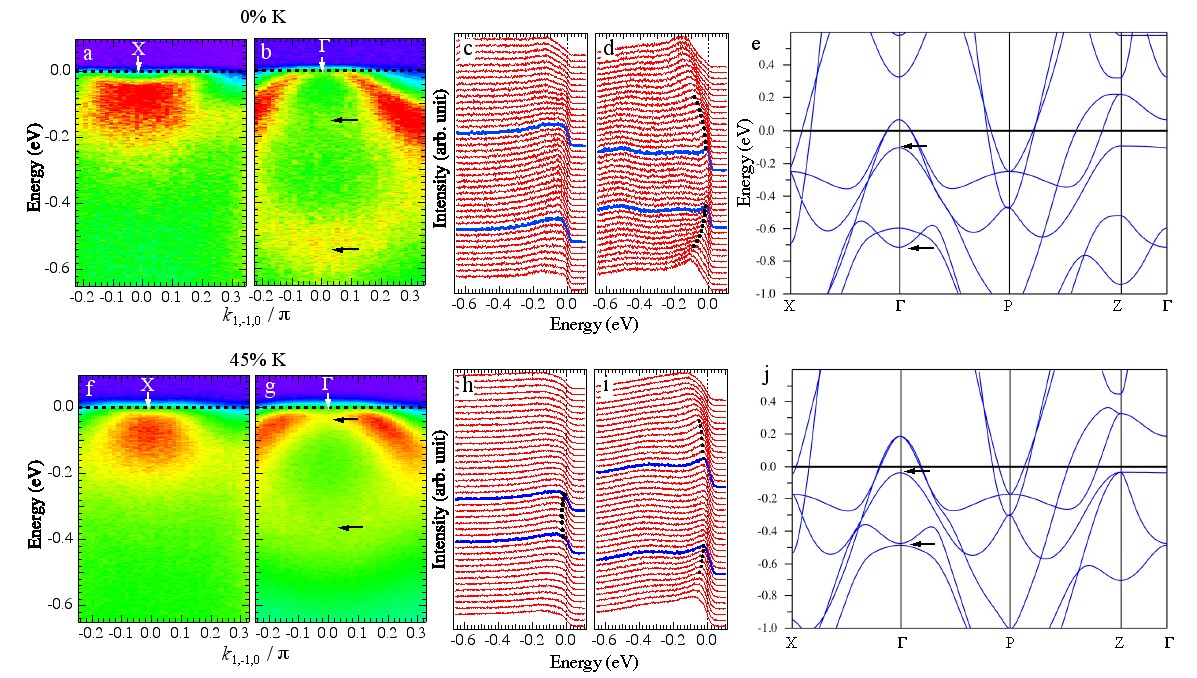}
\caption{(color online)
Experimental band dispersions and
corresponding energy distribution curves (EDCs) along selected cuts,
compared with theoretical band structure.
Panels (a)-(d) present the data of undoped BaFe$_2$As$_2$, while panels (e)-(h) present the
data of Ba$_{1-x}$K$_x$Fe$_2$As$_2$ with nominal $x = 0.45$.
a) ARPES intensity map in X region for undoped sample.
b) ARPES intensity map in $\Gamma$ region for undoped sample.
c,d) EDC's for same cuts as in (a) and (b). Curves at Fermi momentum are marked blue.
e) Results of FLAPW band calculation for parent compound BaFe$_2$As$_2$.
f) ARPES intensity map in X region for doped sample x=0.45.
g) ARPES intensity map in $\Gamma$ region for doped sample x=0.45.
h,i) EDC's for same cuts as in (f) and (g). Curves at Fermi momentum are marked blue.
e) Results of FLAPW band calculation for doped Ba$_{1-x}$K$_x$Fe$_2$As$_2$ with x=0.45.
} \label{fig3}
\end{figure*}

The shape of the Fermi surface (FS) is normally illustrated by
plotting the photoelectron intensity at the chemical
potential\cite{JOELFS,HELENFS}. In Fig. 1a and b we plot this
quantity integrated within 20 meV energy range for undoped and
potassium-doped samples. The Fermi surface of undoped BaFe$_2$As$_2$
consists of a smaller circular-shaped hole pocket centered at
$\Gamma$ and larger electron pockets at X points with also a
circular shape, whose spectrum at two corners in the $k_{1,-1,0}$
direction is somewhat more intense. In the potassium doped samples
(Fig. 1b) the $\Gamma$ pocket becomes larger and the X pocket
shrinks, which is consistent with hole doping of carriers. We note
that the X-pockets in doped samples have a characteristic
``starrish" shape, in reasonable agreement with FLAPW calculations
shown in Figs. 1c-e. The variation of intensity around each contour
of both $\Gamma$ and X pockets is due to photoemission matrix
elements.

In Fig. 2 we show how the shape of both the $\Gamma$ and X bands
evolves with binding energy and potassium doping. With increasing
binding energy, the contour of the $\Gamma$ pocket becomes larger -
consistent with its hole-like topology, while the X pocket becomes
smaller - consistent with its electron-like topology. One notable
fact shown in Fig. 2 is that the size of the pockets at a binding
energy 0.04eV in the parent compound is similar to that of the
potassium doped compound at the Fermi level. In other words, based
on a rigid-band shifting scenario, we could approximately say that
the potassium hole doping lowers the chemical potential by $\sim$40
meV. We note also that with potassium doping the bottom of the X
band is located in very close proximity to the chemical potential.

In figure 3 we plot the experimental band dispersion data
perpendicular to the $\Gamma$-X direction along cuts through the X
and $\Gamma$ points (i.e. along the $k_{1,-1,0}$ direction). For
both the $x = 0$ and $x = 0.45$ doping levels, the hole pockets at
$\Gamma$ and the electron pockets at X are shown (see also Fig. 2)
and are in general agreement with band calculations. Here we can
examine the relative size of the hole pockets and the electron
pockets in more detail by studying the Fermi crossing momenta
($k_{\textrm{F}}$s) marked by blue curves in Figs. 3c, 3d, 3h and
3i. These $k_{\textrm{F}}$s were determined from the most intense
points in the momentum distribution curves (MDCs) at the Fermi
level. The number of \textit{k}-curves between two $k_{\textrm{F}}$s
is proportional to the actual size of each pocket. 

The calculated Fermi surface (Fig. 3e) of BaFe2As2 and LaFeAsO compounds are very similar \cite{Singh,Nekrasov}. However, details of the bandstructure around Fermi energy were sensitive to this parameter, as previously reported \cite{Mazin}. With the change of As position from experimental one to the correspondent of energy minima the shape of Fermi surface sheet corresponded to holes pocket, shown by green color on Fig. 1c, changes from cylinder without dispersion along $k_z$-direction \cite{Nekrasov} to modulated cylinder with strong dispersion along $k_z$. 
The band structure for the doped material performed using virtual crystal method is shown in Fig. 3j. Clearly, in the undoped parent compound both in experimental data as well as calculation the X-pocket is larger than the $\Gamma$
pocket, whereas in the $x = 0.45$ potassium-doped samples, the
opposite is the case. This effect is consistent with the idea of
rigid-band shifting, namely, potassium doping lowers the chemical
potential of the parent compound, while the shapes of the bands are
left unchanged. A second notable doping dependent feature is the
energy shift of two fully occupied bands, marked by black arrows in
Figs. 3b, 3e, 3g and 3j, where Fig. 3e and 3j plot the calculated
band structure of the undoped BaFe$_2$As$_2$ and potassium-doped
Ba$_{1-x}$K$_x$Fe$_2$As$_2$ with $x = 0.45$, respectively. These
arrows point to similar characteristic features in experimental data
and calculations and how they change upon doping. On potassium
doping, the upper band shifts to lower binding energy by $\sim$130
meV, while the lower band shifts to lower binding energy by
$\sim$180 meV. This fact is in qualitative agreement with
calculations. A third feature notable in Fig. 3 is the missing of
bilayer splitting in the measured data. This fact also agrees well
with the FLAPW calculations, where the two bands constructing the
Fermi surface are highly degenerated in both the undoped and K-doped
systems. All these doping dependent features point to the conclusion
that the FLAPW approximation is valid in both undoped and hole-doped
iron arsenic superconductors.

Despite the fact that our potassium-doped samples display bulk
superconductivity, we did not detect a superconducting gap in the
ARPES measurement down to 12K. This may be due to loss of the
potassium from the surface in the ultra-high vacuum (UHV)
environment or possibly a variation in potassium doping between
different layers, which suggests a strong dependence of the critical
temperature on doping. Our data also indicates that upon further
doping to a point where the material becomes superconducting, the
bottom of the X pocket will be in very close proximity to the
chemical potential. Such an increase of the density of states (DOS)
may be relevant to the emergence of superconductivity in these
materials.

In conclusion we have determined the evolution of the Fermi surface
and band dispersion for Ba$_{1-x}$K$_x$Fe$_2$As$_2$ for $x = 0$ and
$x = 0.45$. We find in the undoped samples, the Fermi surface
consists of a smaller hole pocket centered at $\Gamma$ (0,0,0) and a
larger electron pocket located at each of X points. This is in
general agreement with band calculations. Upon doping, the $\Gamma$
hole pocket increases in size and the X pocket contracts, which is
consistent with hole doping. The conduction bands shift in energy
with doping in a similar fashion, in accordance with a rigid band
shift scheme. Our data indicates that upon further doping to a point
where the material becomes superconducting, the bottom of the X
pocket will be in very close proximity to the chemical potential.
The consequent increase in the DOS may be relevant for the emergence
of superconductivity in these materials.

We are grateful for useful discussions with J\"{o}rg Schmalian. We
thank Helen Fretwell for useful remarks and corrections. Work at
Ames Laboratory was supported by the Department of Energy - Basic
Energy Sciences under Contract No. DE-AC02-07CH11358. ALS is
operated by the US DOE under Contract No. DE-AC03-76SF00098.
Brookhaven National Laboratory is supported by US DOE under Contract
No. DE-AC02-98CH10886. AFSS thanks LPEM for financial support.

%%%%%%%%%%%%%%%%%%%%%%

\end{document}